\documentclass[sigconf]{acmart}
\usepackage{siunitx}
\usepackage{graphicx}
\usepackage{float}
\usepackage{caption, comment}
\usepackage{mathtools, multirow}
\usepackage{booktabs}
\usepackage{array}
\usepackage{makecell}
\usepackage[caption=false]{subfig}
\usepackage{color}
\usepackage{bm}

\renewcommand\footnotetextcopyrightpermission[1]{} 
\AtBeginDocument{%
  \providecommand\BibTeX{{%
    \normalfont B\kern-0.5em{\scshape i\kern-0.25em b}\kern-0.8em\TeX}}}

\usepackage{tikz}
\newcommand\encircle[1]{%
\tikz[baseline=(char.base)] 
  \node (char) [draw, scale=0.75, shape=circle, inner sep=0, fill=black, text=white, minimum size=0em] {\strut #1};}

\copyrightyear{2026}
\acmYear{2026}
\setcopyright{rightsretained}
\acmConference[ICCAD '26]{ACM/IEEE 2026 INTERNATIONAL
CONFERENCE ON
COMPUTER-AIDED
DESIGN}{November 8--12, 2026}{San Jose, CA, USA}
\acmBooktitle{61st ACM/IEEE Design Automation Conference (DAC '24), June 23--27, 2026, San Francisco, CA, USA}
\acmDOI{10.1145/3649329.3657359}
\acmISBN{979-8-4007-0601-1/24/06}

\begin{document}

\title[Opto-ViT-v2]{Opto-ViT-v2: Noise-Resilient On-Chip Fine-Tuning for Photonic Near-Sensor Vision Transformer Accelerators}

\author{Xuming Chen}
\email{xuming.chen@case.edu}
\affiliation{
  \institution{Case Western Reserve University}
  \country{USA}
}

\author{Deniz Najafi}
\email{dn339@njit.edu}
\affiliation{
  \institution{New Jersey Institute of Technology}
  \country{USA}
}

\author{Mehrdad Morsali}
\email{mm2772@njit.edu}
\affiliation{
  \institution{New Jersey Institute of Technology}
  \country{USA}
}

\author{Chengwei Zhou}
\email{chengwei.zhou@case.edu}
\affiliation{
  \institution{Case Western Reserve University}
  \country{USA}
}

\author{Zahra Ghanaatianjobzari}
\email{Zahra.Ghanaatian@colostate.edu}
\affiliation{
  \institution{Colorado State University}
  \country{USA}
}

\author{Mahdi Nikdast}
\email{mahdi.nikdast@colostate.edu}
\affiliation{
  \institution{Colorado State University}
  \country{USA}
}

\author{Shaahin Angizi}
\email{shaahin.angizi@njit.edu}
\affiliation{
  \institution{New Jersey Institute of Technology}
  \country{USA}
}

\author{Gourav Datta}
\email{gourav.datta@case.edu}
\affiliation{
  \institution{Case Western Reserve University}
  \country{USA}
}
\renewcommand{\shortauthors}{X. Chen et al.}

\begin{abstract}
Silicon-photonic (SiPh) accelerators have emerged as a
promising platform for Vision Transformer (ViT) inference,
achieving high throughput and energy efficiency by performing
matrix multiplications on microring-resonator (MRR) banks.
Extending these platforms to support on-chip fine-tuning,
however, has remained an open problem: backpropagation demands
large activation storage that exceeds typical on-chip buffer
capacity, weight write-back to MRR banks after each gradient
step is prohibitively slow, and device-level noise corrupts
both forward and backward computation.
We present \textbf{Opto-ViT-v2}, the first framework that
enables parameter-efficient fine-tuning (PEFT) on a near-sensor
SiPh ViT accelerator. We propose a tensorized low-rank
decomposition of the weight increments that separates pretrained
weights (executed on the optical cores) from a small set of
trainable factors (as few as 8K parameters for ViT-Base) that
reside and are updated entirely in the electronic domain. This
decomposition dramatically reduces both the activation storage
required during backpropagation and the number of weight
write-back operations, making on-chip training feasible within
the memory and tuning constraints of the photonic architecture.
We further introduce a \emph{gradient-accumulated sparse head}
that freezes low-importance classifier weights via one-shot
top-$k$ gradient masking, reducing the head training budget
by $\sim$40\%.
To evaluate training under realistic hardware conditions, we
develop the first system-level noise model for photonic
on-chip training, characterizing how MRR crosstalk, thermal
drift, and laser amplitude noise propagate through both
forward and backward passes. Calibrated against $>$200
fabricated MRR devices, this model enables evaluation of PEFT
methods under photonic noise and reveals that low-rank
factor updates exhibit greater robustness than full
fine-tuning or per-layer low-rank adaptation under identical
noise conditions.
Evaluated on VTAB-1K (19~tasks) and FGVC few-shot benchmarks
using a full-stack device-to-application co-design framework,
Opto-ViT-v2 recovers within 0.3-0.8\% of clean-software accuracy
under measured photonic noise, achieving $>$100~KFPS/W
throughput and enabling, for the first time, on-chip domain
adaptation at the photonic sensor edge.
\end{abstract}

\maketitle
 
\section{Introduction}
\label{sec:intro}
 
Vision Transformers (ViTs)~\cite{dosovitskiy2021image} have
become the dominant backbone for tasks spanning image
classification, object detection, and video
understanding, owing to their scalable self-attention
mechanism and strong transfer-learning capability. The standard
deployment paradigm pretrains a ViT on a large-scale dataset
(e.g., ImageNet-21K) and then fully fine-tunes it on each
downstream task. This paradigm, however, poses two compounding
challenges for resource-constrained edge systems. First, the
dense matrix multiplications in multi-head self-attention (MHSA)
and feed-forward network (FFN) blocks are computationally
intensive, demanding accelerator-class hardware even for
inference. Second, adapting the model to new domains requires
updating tens of millions of parameters, a process that is
memory- and energy-prohibitive on embedded platforms.
 
Silicon-photonic (SiPh) accelerators offer an attractive path
toward addressing the inference challenge. By encoding weights
onto microring-resonator (MRR) banks and performing
multiply-accumulate (MAC) operations via wavelength-division
multiplexing (WDM), architectures such as
CrossLight~\cite{sunny2021crosslight},
Lightator~\cite{morsali2024lightator}, and
Opto-ViT~\cite{morsali2025optovit} have demonstrated
inference throughput exceeding 100~KFPS/W with significant
energy savings over digital baselines. These systems stream
pretrained weight matrices onto MRR arrays layer by layer and
execute the ViT inference pipeline through a hybrid
electronic-photonic dataflow.
 
The second challenge, on-chip fine-tuning, remains unsolved
for photonic platforms. Three fundamental constraints make
backpropa-gation-based training difficult on SiPh hardware.
(i)~\emph{Activation storage:}
Backpropagation requires intermediate activations from the
forward pass to compute parameter gradients during the
backward pass. 
Storing them
on-chip alongside the weight parameters far exceeds the
buffer memory capacity of current photonic architectures;
recomputing them from scratch for each layer during the
backward pass, while possible, multiplies the number of
optical core invocations and is prohibitively expensive.
(ii)~\emph{Weight write-back:}
In standard fine-tuning, updated weights must be written back to the MRR banks after each gradient step. Repeated over many mini-batches and epochs, this write-back process dominates latency and energy, fundamentally limiting training throughput.
(iii)~\emph{Device noise:}
Photonic MAC operations are subject to fabrication
variability, thermal crosstalk between adjacent MRRs, and
laser relative intensity noise, all of which corrupt both
forward activations and backward gradient signals, degrading
training convergence.
 
Parameter-efficient fine-tuning (PEFT) methods offer a path
forward by dramatically reducing the number of trainable
parameters. Methods such as LoRA~\cite{hu2022lora}, visual
prompt tuning (VPT)~\cite{jia2022visual}, and
adapters~\cite{houlsby2019parameter} have shown that updating
fewer than 1\% of a ViT's parameters can match or exceed full
fine-tuning accuracy on diverse downstream
tasks~\cite{jie2023fact,chen2022adaptformer,zhang2022noah}.
Recent work on tensorized decomposition of weight
increments~\cite{jie2023fact} has pushed this further,
achieving competitive accuracy with as few as 0.01\% of trainable
parameters by exploiting both
intra-layer and cross-layer redundancy. However, no prior
work has investigated how to map PEFT training onto photonic
hardware, or whether the reduced parameter footprint
translates into concrete architectural benefits under the
activation storage, write-back, and noise constraints of SiPh
accelerators.
 
In this paper, we present \textbf{Opto-ViT-v2}, \emph{the first
framework enabling on-chip PEFT on a near-sensor
silicon-photonic ViT accelerator, targeting real-world vision tasks.} We show that a tensorized
low-rank decomposition of weight increments is not merely
parameter-efficient in the software sense, but directly
addresses all three photonic training bottlenecks: the small
trainable factors require orders of magnitude less activation
caching during backpropagation, eliminate MRR weight
write-back entirely (since pretrained weights remain static on
the optical cores), and confine all parameter updates to the
electronic domain where they are immune to photonic noise.
Our contributions are as follows.
 
\noindent\textbf{(1) First on-chip photonic PEFT via hardware–algorithm co-design.}
We develop a tensorized low-rank decomposition strategy co-designed with the Opto-ViT-v2 framework: pretrained weights $\mathcal{W}_0$ are streamed onto MRR banks during forward and backward passes, while a compact set of trainable factors resides entirely in the electronic processing unit. By keeping $\mathcal{W}_0$ fixed, we eliminate MRR retuning and costly weight write-backs. Factor gradients are computed via small ($d{\times}r$, $r{\times}r$) operations without forming full $d{\times}d$ gradients, significantly reducing compute and memory overhead. We further design the complete forward–backward dataflow, including optical scheduling, activation compression leveraging low-rank structure, and gradient checkpointing aligned with the Opto-ViT-v2 memory hierarchy.
 
\noindent\textbf{(2) Gradient-accumulated sparse head.}
While tensorized decomposition significantly reduces the backbone’s trainable parameters, the task-specific classification head comes to dominate the budget, accounting for over 90\% of the total trainable parameters. Applying low-rank factorization to the head degrades accuracy because the classifier must produce sharp, class-discriminative logits that aggressive rank reduction
blurs. We propose a complementary approach that accumulates gradients over multiple mini-batches to derive robust importance estimates, and then retains only the top-$k$ parameters (by gradient magnitude) as trainable while freezing the rest. This \emph{gradient-accumulated sparse head} reduces the overall training budget by $\sim$40\% with negligible accuracy loss. 
 
\noindent\textbf{(3) Full-stack evaluation with photonic
training noise model.}
The first system-level noise model for photonic on-chip training is designed, capturing how MRR crosstalk, thermal drift, and laser amplitude noise propagate through both forward and backward passes. Calibrated against measurements from $>$200 fabricated MRR devices, the model enables realistic evaluation of PEFT methods under photonic noise. Using a full-stack device-to-application framework, we show that Opto-ViT-v2 recovers within 0.3--0.8\% of noise-free accuracy while achieving $>$100 KFPS/W, and that low-rank factor updates exhibit improved robustness compared to full fine-tuning and per-layer adaptation.

\section{Background \& Related Work}
\label{sec:background}

\noindent\textbf{Vision Transformer Architecture}: As shown in Fig. 1, A Vision Transformer (ViT)~\cite{dosovitskiy2021image} partitions
an input image into non-overlapping patches, projects each patch
into a $d$-dimensional embedding, and processes the resulting
sequence through $L$ encoder blocks. Each block consists of a
multi-head self-attention (MHSA) module followed by a feed-forward
network (FFN), with layer normalization and residual connections
around each sub-layer. In MHSA, the input $\mathbf{X} \in \mathbb{R}^{N \times d}$
(where $N$ is the number of patches) is projected into queries,
keys, and values via weight matrices
$\mathbf{W}_Q, \mathbf{W}_K, \mathbf{W}_V \in
\mathbb{R}^{d \times d}$, which are further divided into $N_h$
heads of dimension $d_k = d / N_h$. The self-attention output
for head $i$ is
\begin{equation}
\text{Head}_i(\mathbf{X}) =
  \text{softmax}\!\left(
    \frac{\mathbf{X}\mathbf{W}_Q^{(i)}
          {\mathbf{W}_K^{(i)}}^\top \mathbf{X}^\top}
         {\sqrt{d_k}}
  \right)
  \mathbf{X}\mathbf{W}_V^{(i)}{\mathbf{W}_O^{(i)}}^\top
\label{eq:mhsa}
\end{equation}
and the outputs of all heads are concatenated and linearly
projected. The FFN applies two fully connected layers with a
GELU nonlinearity:
$\text{FFN}(\mathbf{X}) =
  \text{GELU}(\mathbf{X}\mathbf{W}_\text{up})
  \mathbf{W}_\text{down}$,
where $\mathbf{W}_\text{up} \in \mathbb{R}^{d \times 4d}$ and
$\mathbf{W}_\text{down} \in \mathbb{R}^{4d \times d}$. For ViT-Base/16, $L{=}12$, $d{=}768$, $N_h{=}12$, and
$d_k{=}64$, yielding approximately 86M total parameters. Each
encoder block contains four $d {\times} d$ matrices in the
MHSA module ($\mathbf{W}_Q$, $\mathbf{W}_K$, $\mathbf{W}_V$,
$\mathbf{W}_O$) and two matrices in the FFN
($\mathbf{W}_\text{up}$, $\mathbf{W}_\text{down}$), with the
FFN matrices subdivided into $d {\times} d$ blocks as described
in~\cite{jie2023fact}. This structural uniformity across layers
is important for the tensorized decomposition we employ in
Section~\ref{sec:method}. 

\begin{figure} 
\centering
\vspace{-4mm}
\includegraphics[width=0.8\linewidth,]{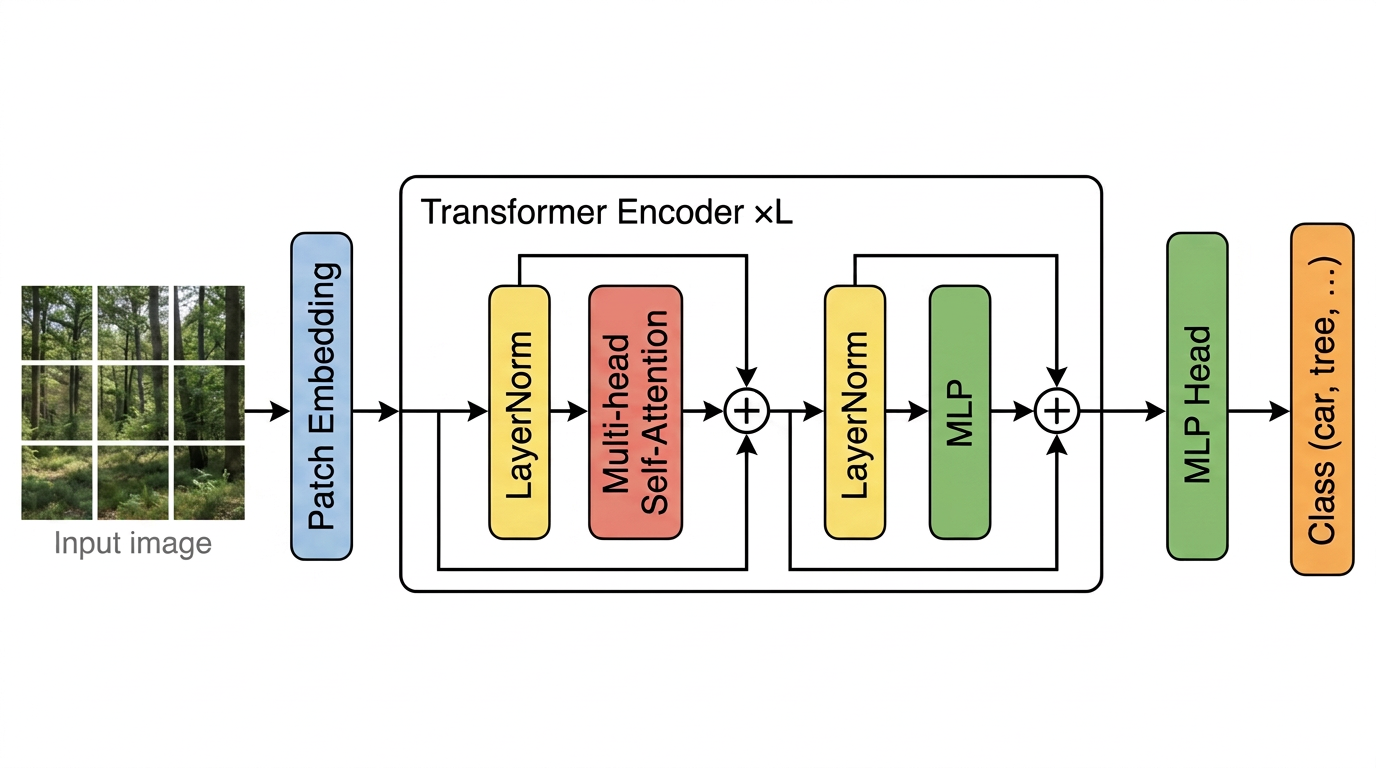}
\vspace{-3.2em}
\caption{\footnotesize Architecture overview of a standard ViT encoder that processes patch embeddings through $L$ MHSA–FFN layers with layer normalization and residuals.} 
\vspace{-1.5em}
\label{mr}
\end{figure}


\noindent\textbf{Silicon-Photonic ViT Acceleration}: Silicon-photonic (SiPh) accelerators perform MACs in the optical domain by encoding inputs as light amplitudes and weights as microring resonator (MRR) transmission coefficients. With wavelength-division multiplexing (WDM), multiple multiplications occur in parallel across wavelengths, and balanced photodetectors accumulate results. MRR weights are programmed via thermo- or electro-optic tuning, shifting resonance to modulate overlapping optical signals. Prior SiPh accelerators target specific DNN settings: LightBulb~\cite{zokaee2020lightbulb} for binarized CNNs (limited by ADC power), HolyLight~\cite{liu2019holylight} and CrossLight~\cite{sunny2021crosslight} for reduced-precision convolutions, ROBIN~\cite{sunny2021robin} for robustness, Lightator~\cite{morsali2024lightator} for near-sensor processing, TRON~\cite{afifi2023tron} for transformer blocks, and Opto-ViT~\cite{morsali2025optovit} for hybrid photonic–electronic ViTs with WDM parallelism and pruning. However, none support on-chip training.

\begin{figure}[t] 
\centering
\includegraphics [width=0.99\linewidth,]{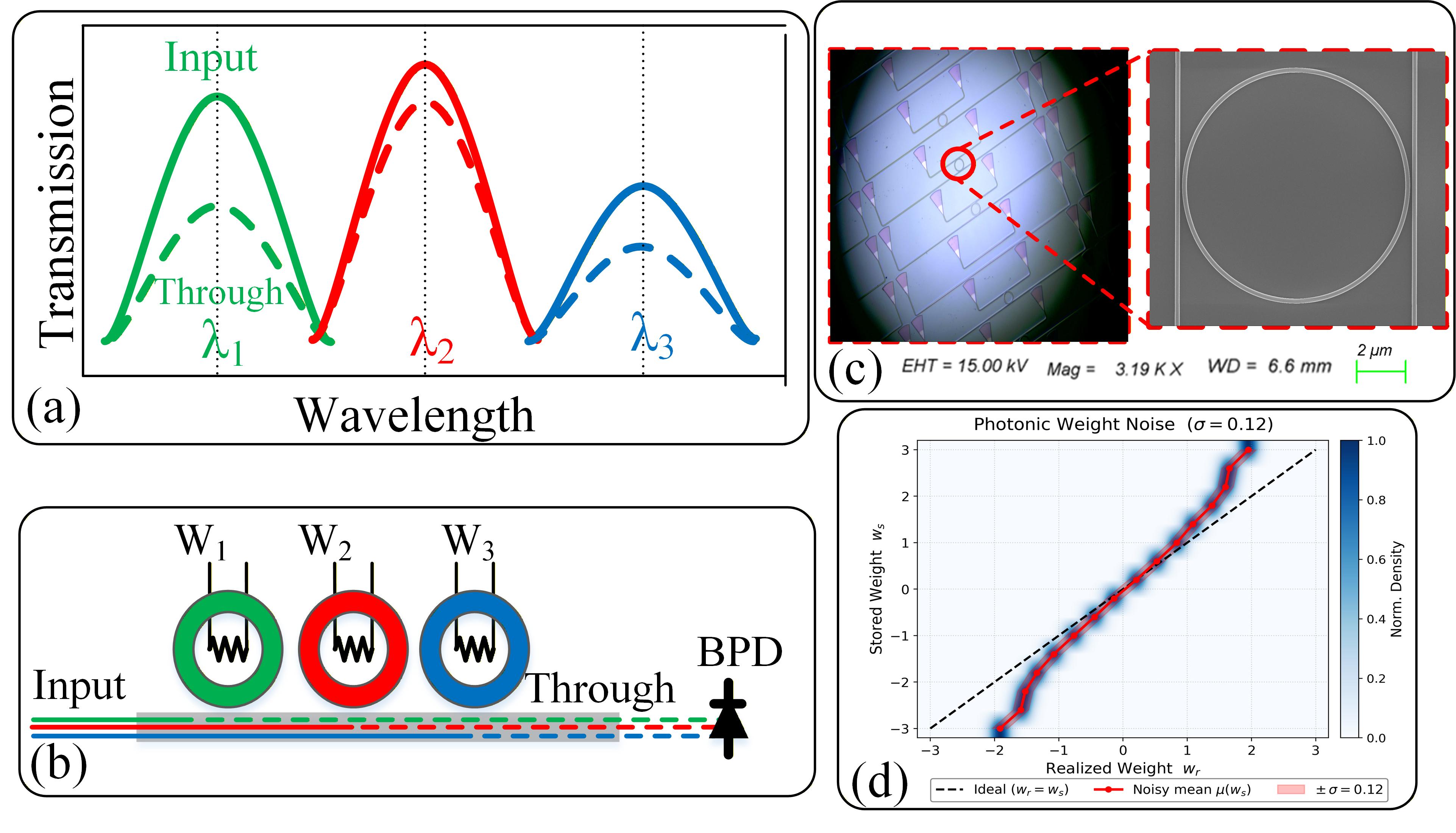}
\vspace{-1.8em}
\caption{\footnotesize (a) Spectra at the MRR input and through ports after parameter imprinting via tuning. Adjusting the resonant wavelength causes part of the signal to couple into the ring while the rest passes to the through port, encoding the parameter in the output.
(b) Multiple MRRs in one arm imprint weight values across different wavelengths of the input signal.
(c) Fabricated SiPh IC with over 200 identical MRR cell layouts with SEM image.
(d) Photonic weight noise.} 
\vspace{-1.5em}
\label{device}
\end{figure}

\noindent\textbf{Parameter-Efficient Fine-Tuning for ViTs}: Parameter-efficient transfer learning (PETL) adapts pretrained models by updating only a small subset of parameters, reducing storage and compute. {Adapter methods} insert lightweight bottlenecks~\cite{houlsby2019parameter,pfeiffer2021adapterfusion,chen2022adaptformer}, while {prompt-based methods} prepend learnable tokens~\cite{jia2022visual,zhang2022noah}. {Low-rank methods} like LoRA~\cite{hu2022lora} decompose each weight update as $\Delta\mathbf{W} = s\mathbf{A}\mathbf{B}$, but operate independently per layer. {Tensorized methods} address this by jointly decomposing all weights. FacT~\cite{jie2023fact} stacks all ViT weight matrices into a tensor $\mathcal{W} \in \mathbb{R}^{12L \times d \times d}$, where {$12L$ denotes the total number of $d{\times}d$ matrices across $L$ layers} (4 from MHSA and 8 from FFN per layer). The update $\Delta\mathcal{W}$ is then factorized (e.g., Tensor-Train), sharing factors across layers. At rank $r{=}4$, FacT-TT uses only $\sim$8K parameters ($\sim$0.01\% of ViT-Base) while achieving competitive accuracy.
 
 
\section{Proposed Method: Opto-ViT-v2}
\label{sec:method}

\begin{figure}
\centering
\includegraphics [width=1.01\linewidth,]{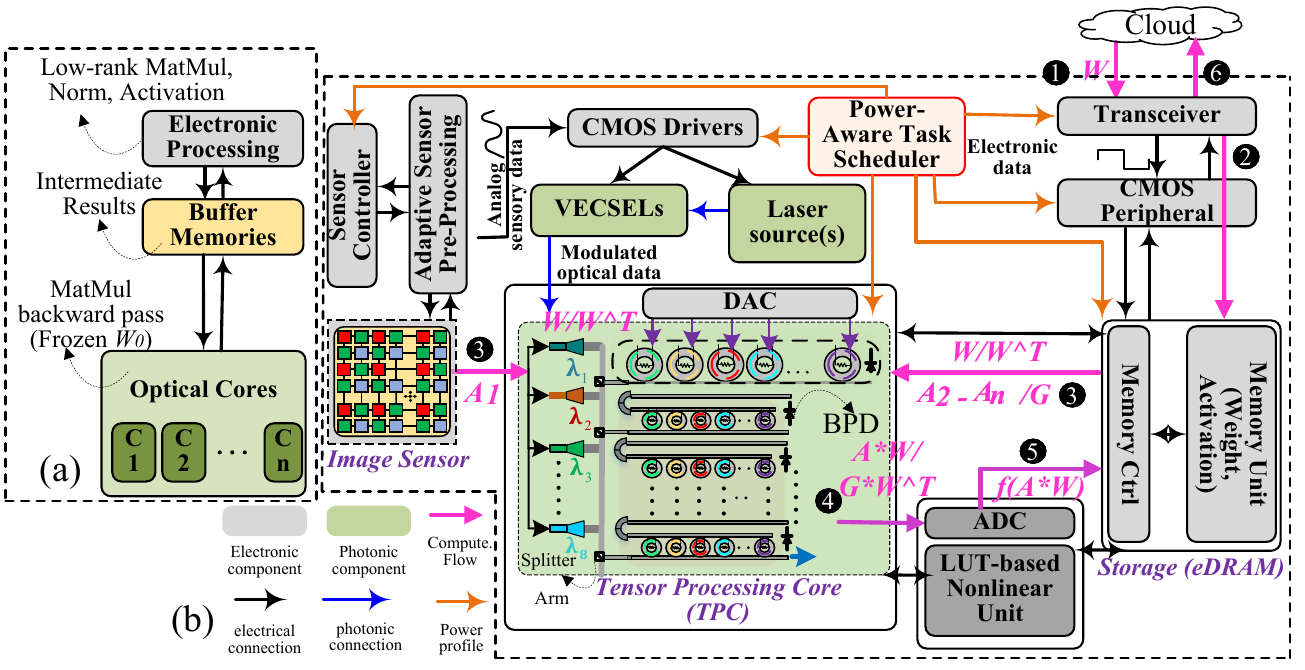}
\vspace{-2em}
\caption{(a) Opto-ViT-v2 operations, and (b) photonic architecture and corresponding interface signals for Opto-ViT-v2.} 
\vspace{-1.8em}
\label{arch}
\end{figure}

Fig. \ref{arch} provides an overview of the proposed Opto-ViT-v2 node architecture, highlighting its electronic and photonic building blocks, which are \textit{co-designed and co-optimized} to realize a standalone, reconfigurable, and energy-aware optical vision sensor that seamlessly integrates sensing and processing. \textit{The design intrinsically supports granularity-configurable, low-bit-width implementations of MHSA, patch embedding and linear layers in ViTs. This enables flexible trade-offs between power consumption and accuracy.} Opto-ViT-v2 consists of the following steps.

In step~\encircle{1}, the weights are received from the cloud via an RF transceiver and written to on-chip
memory~\encircle{2}. In
step~\encircle{3}, the frozen weights are streamed layer by
layer onto the Tensor Processing Core (TPC) via the memory
controller. The TPC comprises 64 waveguide arms, each
supporting 32 wavelength channels (aligned with $d_k$), with
MRRs performing parallel multiplication and BPDs accumulating
results. Input activations are captured by a power-aware CMOS
image sensor and modulated directly onto VCSEL-generated
optical signals, avoiding the higher cost of MRR-based input
encoding. In step~\encircle{4}, MAC operations execute in
parallel on the TPC, with results fed to a programmable
LUT-based nonlinear unit for activation functions and their
derivatives. 
In step~\encircle{5},
TPC outputs are digitized via ADCs and written to the
activation memory, with compressed intermediates cached for the
backward pass. In step~\encircle{6}, trained factors can be
absorbed into the weights for zero-overhead inference, and
results are transmitted to the cloud if required.

We present three contributions that together form a complete
on-chip fine-tuning framework: Section~\ref{sec:method_hwmap}
maps the tensorized low-rank decomposition onto the TPC and
electronic unit, deriving the forward and backward dataflows
that keep $\mathcal{W}_0$ frozen on the optical cores while
confining all trainable computation to the electronic domain.
Section~\ref{sec:method_head} introduces a
gradient-accumulated sparse head that reduces the training budget by $\sim$40\%.
Section~\ref{sec:method_noise} develops the first system-level
noise model for photonic training, characterizing how MRR
device noise in both optical passes propagates into the factor
gradients computed electronically in
Section~\ref{sec:method_hwmap}.
 
 \begin{figure}[t] 
 \centering
\includegraphics[width=0.38\textwidth]{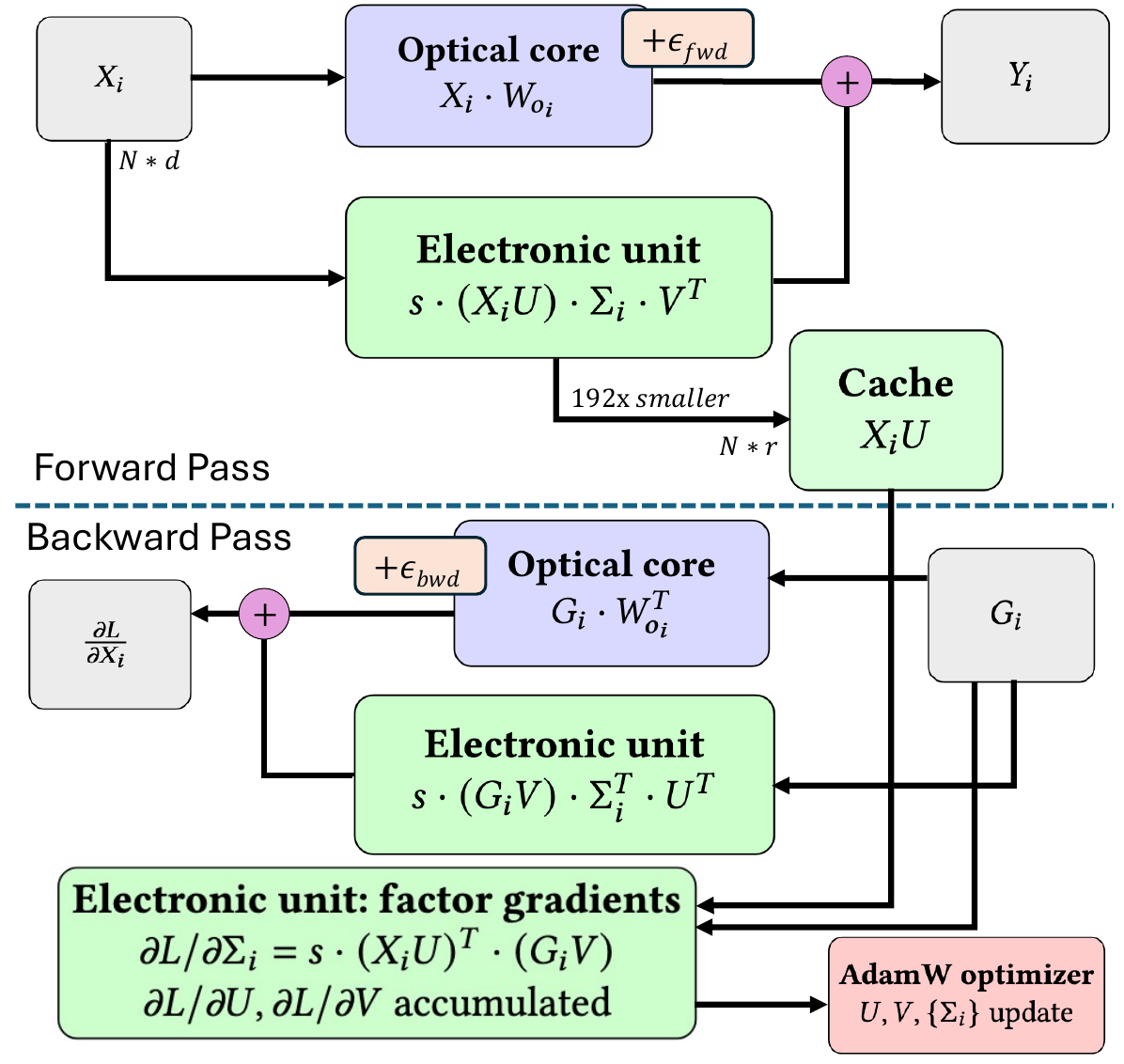} 
\caption{Opto-ViT-v2 per-layer training dataflow. Optical cores (blue) handle the heavy $W_0$ multiplications in both passes with photonic noise (amber); the electronic unit (green) computes the lightweight factor chain and all factor gradients using the compressed cached intermediate $X_iU$ (192× SRAM reduction). $W_0$ is never retuned}
\label{ev2}
\end{figure}

\subsection{Hardware Mapping of Tensorized Low-Rank Training}
\label{sec:method_hwmap}
 
Training neural networks on analog or mixed-signal hardware is
fundamentally challenging. Prior work on analog in-memory
computing (AIMC)~\cite{rasch2024fast} and ReRAM-based
transformer accelerators~\cite{yang2020retransformer} has
identified weight write-back overhead, limited update precision,
and intermediate data buffering as primary obstacles. In SiPh
accelerators, the analogous bottlenecks are MRR retuning
latency, buffer memory capacity, and photonic MAC noise. These constraints make standard backpropagation with full weight updates infeasible on the Opto-ViT-v2 platform.
 
As shown in Fig. 4, our approach is to decompose the fine-tuning weight update into
a heavy frozen component executed on the optical cores and a
light trainable component confined to the electronic domain.
We adopt the tensorized Tensor-Train
decomposition~\cite{jie2023fact},
in which the weight of the $i$-th sub-layer during fine-tuning
is expressed as $\mathbf{W}_i = \mathbf{W}_{0,i}
  + s \cdot \mathbf{U}\,\bm{\Sigma}_i\,\mathbf{V}^\top
\label{eq:weight_decomp}$, where $\mathbf{W}_{0,i} \in \mathbb{R}^{d \times d}$ is the
frozen pretrained weight,
$\mathbf{U} \in \mathbb{R}^{d \times r}$ and
$\mathbf{V} \in \mathbb{R}^{d \times r}$ are global factors
shared across all $12L$ sub-layers,
$\bm{\Sigma}_i \in \mathbb{R}^{r \times r}$ is specific to
sub-layer~$i$, $s$ is a scaling hyperparameter, and $r \ll d$.
Only $\mathbf{U}$, $\mathbf{V}$, and
$\{\bm{\Sigma}_i\}_{i=1}^{12L}$ are trainable. For ViT-Base
($L{=}12$, $d{=}768$) at rank $r{=}4$, the total trainable
parameter count is
$2dr + 12Lr^2 = 2 \cdot 768 \cdot 4 + 144 \cdot 16 = 8{,}448$,
occupying fewer than 8.5~KB at 8-bit precision.
 
We now derive how this decomposition maps onto the Opto-ViT-v2
hardware for both the forward and backward passes, and present
the five-core scheduling strategy.

\subsubsection{Forward Pass}
\label{sec:method_fwd}
 
The forward pass through sub-layer~$i$ computes the output
$\mathbf{Y}_i$ from input
$\mathbf{X}_i \in \mathbb{R}^{N \times d}$:
\begin{equation}
\mathbf{Y}_i
  = \underbrace{\mathbf{X}_i\,\mathbf{W}_{0,i}}_{\text{optical}}
  + s \cdot
    \underbrace{(\mathbf{X}_i\,\mathbf{U})\,
    \bm{\Sigma}_i\,\mathbf{V}^\top}_{\text{electronic}}
\label{eq:fwd}
\end{equation}
The first term is a $N {\times} d$ by $d {\times} d$ matrix
multiplication requiring $Nd^2$ MACs. This is the dominant
computational cost and is executed on the optical cores, with
$\mathbf{W}_{0,i}$ streamed onto the MRR banks layer by layer
in 32-element chunks following the partitioning strategy
described in \cite{morsali2025optovit}. Since
$\mathbf{W}_{0,i}$ is never modified, the MRR banks serve as
a static weight store throughout the entire fine-tuning
process, and no retuning or write-back is required. The second term decomposes into a chain of three small matrix
multiplications executed on the electronic processing unit:
(a)~$\mathbf{X}_i \mathbf{U}$:
  $N {\times} d$ by $d {\times} r \to N {\times} r$
  ($Ndr$ MACs);
(b)~$(\mathbf{X}_i \mathbf{U})\,\bm{\Sigma}_i$:
  $N {\times} r$ by $r {\times} r \to N {\times} r$
  ($Nr^2$ MACs);
(c)~result $\times\,\mathbf{V}^\top$:
  $N {\times} r$ by $r {\times} d \to N {\times} d$
  ($Ndr$ MACs).
For $N{=}196$, $r{=}4$, and $d{=}768$, the electronic factor
chain requires $2Ndr + Nr^2 \approx 1.2$M MACs, compared to
$Nd^2 \approx 115$M MACs for the optical term (<1.1\% overhead), and can run in parallel.
 
\noindent\textbf{Compressed activation caching.}
A key architectural benefit of the tensorized decomposition is
the reduction in activation storage during training.
Standard backpropagation caches the full input
$\mathbf{X}_i \in \mathbb{R}^{N \times d}$ at each sub-layer
for use in the gradient computation. With $12L{=}144$
sub-layers in ViT-Base, the total activation footprint is
$144 \times 196 \times 768 \times 2\text{B} \approx 42$~MB,
far exceeding the buffer memory of Opto-ViT-v2. In our framework, the factor gradient for $\bm{\Sigma}_i$
(Eq.~(\ref{eq:grad_sigma}) below) depends only on the
projected intermediate
$\mathbf{X}_i \mathbf{U} \in \mathbb{R}^{N \times r}$, not
the full $\mathbf{X}_i$. We therefore cache only this small
product during the forward pass. For $r{=}4$, this compresses
per-layer activation storage from $196 \times 768 = 150{,}528$
values to $196 \times 4 = 784$ values, a $\mathbf{192\times}$
\textbf{reduction}. Across all 144 sub-layers, the total
cached footprint drops to
$144 \times 784 \times 2\text{B} \approx 220$~KB, which
comfortably fits in the electronic unit's SRAM.

\subsubsection{Backward Pass}
\label{sec:method_bwd}
 
The backward pass must accomplish two tasks: propagate the
gradient signal backward through the network, and compute the
gradients of the loss with respect to the trainable factors.
 
\noindent\textbf{Gradient propagation.}
Let $\mathbf{G}_i \triangleq
\partial\mathcal{L}/\partial\mathbf{Y}_i
\in \mathbb{R}^{N \times d}$ denote the upstream gradient at
sub-layer~$i$. The gradient with respect to the sub-layer
input is
\begin{equation}
\frac{\partial\mathcal{L}}{\partial\mathbf{X}_i}
  = \underbrace{\mathbf{G}_i\,\mathbf{W}_{0,i}^\top}_{%
      \text{optical}}
  + s \cdot
    \underbrace{(\mathbf{G}_i\,\mathbf{V})\,
    \bm{\Sigma}_i^\top\,\mathbf{U}^\top}_{%
      \text{electronic}}
\label{eq:grad_prop}
\end{equation}
The first term is a dense $N {\times} d$ by $d {\times} d$
multiplication executed on the optical cores. Since
$\mathbf{W}_{0,i}$ is frozen, its transpose
$\mathbf{W}_{0,i}^\top$ is precomputed offline and streamed
onto the MRR banks during the backward pass, incurring only
the standard layer-by-layer streaming latency with no
additional tuning overhead.
The second term is a chain of small electronic multiplications:
$\mathbf{G}_i \mathbf{V}$ ($N {\times} r$), then
$\times\,\bm{\Sigma}_i^\top$ ($N {\times} r$), then
$\times\,\mathbf{U}^\top$ ($N {\times} d$), contributing
negligible overhead relative to the optical term.
 
\noindent\textbf{Factor gradient: per-layer core
$\bm{\Sigma}_i$.}
The gradient with respect to the per-layer factor is computed
entirely in the electronic domain:
\begin{equation}
\frac{\partial\mathcal{L}}{\partial\bm{\Sigma}_i}
  = s \cdot
    \underbrace{(\mathbf{X}_i \mathbf{U})^\top}_{%
      r \times N,\;\text{cached}}
    \underbrace{(\mathbf{G}_i\,\mathbf{V})}_{%
      N \times r}
\label{eq:grad_sigma}
\end{equation}
The left operand
$(\mathbf{X}_i \mathbf{U}) \in \mathbb{R}^{N \times r}$ was
cached during the forward pass. The right operand
$\mathbf{G}_i \mathbf{V} \in \mathbb{R}^{N \times r}$ is a
small electronic multiplication. The result is an
$r {\times} r$ matrix: for $r{=}4$, a $4 {\times} 4$ output
from two $196 {\times} 4$ operands. This is computed
independently for each of the 144 sub-layers.
 
\noindent\textbf{Factor gradient: shared factors $\mathbf{U}$ and $\mathbf{V}$.} The gradient accumulates contributions from all sub-layers: 
\begin{equation} 
\frac{\partial\mathcal{L}}{\partial\mathbf{U}} = s \sum_{i=1}^{12L} \mathbf{X}_i^\top \, \underbrace{\mathbf{G}_i\,\mathbf{V}\, \bm{\Sigma}_i^\top}_{\triangleq\;\mathbf{Z}_i \;\in\;\mathbb{R}^{N \times r}} \ \ \ \ \ \frac{\partial\mathcal{L}}{\partial\mathbf{V}} = s \sum_{i=1}^{12L} \mathbf{G}_i^\top \, \underbrace{(\mathbf{X}_i\,\mathbf{U})\, \bm{\Sigma}_i}_{\in\;\mathbb{R}^{N \times r}}  \label{eq:grad_u} \end{equation} 
Each per-layer term $\mathbf{Z}_i$ reuses the $\mathbf{G}_i \mathbf{V}$ already computed for Eq.~(\ref{eq:grad_sigma}). The product $\mathbf{X}_i^\top \mathbf{Z}_i \in \mathbb{R}^{d \times r}$ requires the full $\mathbf{X}_i \in \mathbb{R}^{N \times d}$, which was not cached (only $\mathbf{X}_i \mathbf{U}$ was stored). During the backward pass, as layers are processed in reverse order, $\mathbf{X}_i$ can be reconstructed by a single forward evaluation of the preceding sub-layer on the optical cores using the stored layer-boundary checkpoint. This adds one extra optical core invocation per layer but avoids storing the full $N {\times} d$ activations for all layers simultaneously. Similarly, for $\mathbf{V}$, $(\mathbf{X}_i \mathbf{U})\bm{\Sigma}_i$ reuses the cached $\mathbf{X}_i \mathbf{U}$ and the small $\bm{\Sigma}_i$. The product $\mathbf{G}_i^\top (\cdot) \in \mathbb{R}^{d \times r}$ is well within the electronic unit's capacity.

\noindent\textbf{Optimizer step.}
After accumulating the factor gradients across all layers, the
AdamW optimizer is applied to $\mathbf{U}$, $\mathbf{V}$, and
$\{\bm{\Sigma}_i\}$ entirely in the electronic domain. The
optimizer state (first and second moments) adds
$2 \times 8{,}448 = 16{,}896$ values, and the total SRAM
footprint for factors plus optimizer state is under 50~KB.

\subsubsection{Five-Core Scheduling}
\label{sec:method_schedule}
 
The Opto-ViT-v2 architecture provides five optical processing
cores (C1--C5), each with 32 WDM wavelength channels across
64 waveguide arms. We propose a \emph{forward-backward
pipelined} scheduling strategy that reuses the same physical
cores for both passes.
 
\noindent\textbf{Forward phase} (layers $l = 1 \to L$).
Each encoder layer is processed following the Opto-ViT-v2 attention pipeline~\cite{morsali2025optovit}. In stage~F1,
cores C1, C2, and C3 are simultaneously tuned with
$\mathbf{W}_Q$, $\mathbf{W}_K^\top / \sqrt{d_k}$, and
$\mathbf{X}^\top$, computing the decomposed attention score
$(\mathbf{Q}\mathbf{W}_K^\top)\mathbf{X}^\top$ without
buffering the intermediate $\mathbf{K}$
(Eq.~(2) in~\cite{morsali2025optovit}). In stage~F2,
the electronic unit computes softmax and C4--C5 process the
value projection and output via $\mathbf{W}_V$ and the
softmax result. The FFN sub-layer is then handled by
streaming $\mathbf{W}_\text{up}$ and $\mathbf{W}_\text{down}$
onto C1--C2 (two cores suffice since the FFN has two
sequential linear layers). Throughout the forward phase, the
electronic unit computes the factor contributions
$s\,(\mathbf{X}_i \mathbf{U})\,\bm{\Sigma}_i\,
\mathbf{V}^\top$ in parallel with the optical computation
and caches $\mathbf{X}_i \mathbf{U}$ for each sub-layer.

 \begin{figure} \vspace{-0.8em}
 \centering
\includegraphics[width=0.40\textwidth]{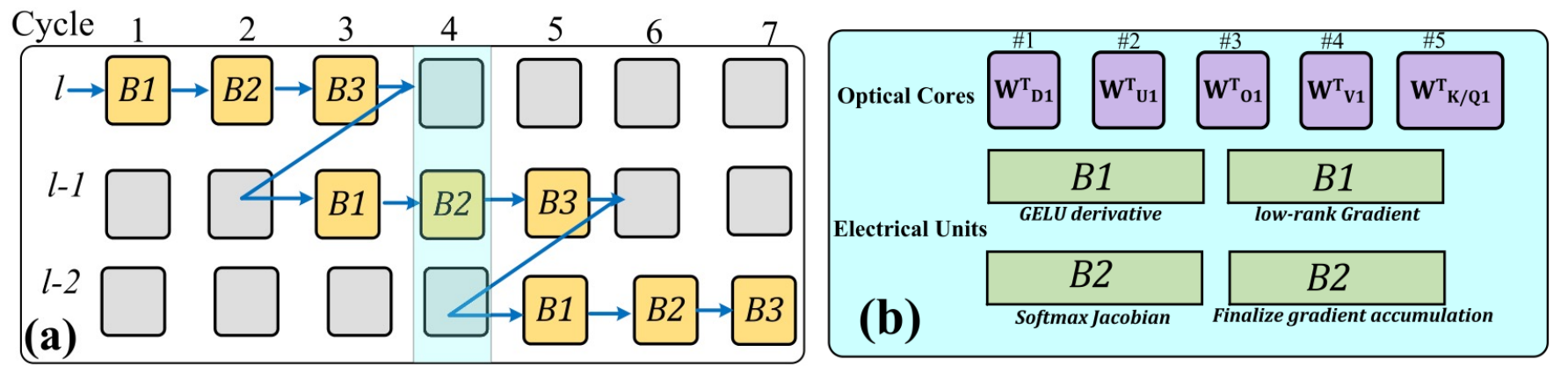} \vspace{-1em}
\caption{\footnotesize (a) Pipelined processing of backward stages of different layers. (b) Optical and Electronic Hardware allocation for backward stage in Cycle 4 }
\vspace{-1.0em}
\label{pipeline}
\end{figure}
 
\noindent\textbf{Backward phase} (layers $l = L \to 1$).
Processing reverses the layer order. For each layer, we define
three stages that overlap optical and electronic operations to
maximize core utilization as shown in Fig. \ref{pipeline}.

\noindent\emph{Stage~B1 (FFN backward):}
Cores C1--C2 are tuned with the precomputed transposes
$\mathbf{W}_\text{down}^\top$ and
$\mathbf{W}_\text{up}^\top$ to propagate the gradient through
the FFN block via Eq.~(\ref{eq:grad_prop}). Concurrently, the
electronic unit computes the GELU derivative and the FFN
factor gradients
(Eqs.~(4)-(6)) using the
cached $\mathbf{X}_i \mathbf{U}$ intermediates and the
upstream gradient $\mathbf{G}_i$.

\noindent\emph{Stage~B2 (MHSA backward):}
Cores C3--C5 are tuned with $\mathbf{W}_O^\top$,
$\mathbf{W}_V^\top$, and the attention backward weights
($\mathbf{W}_Q^\top$, $\mathbf{W}_K^\top$) to propagate the
gradient through the MHSA block. The softmax Jacobian is
computed electronically in parallel. Cores C1--C2, now free
from stage~B1, can assist by handling additional MHSA weight
transposes if needed for the four attention sub-layers.

\noindent\emph{Stage~B3 (overlap with next layer):}
While the optical cores begin streaming weights for layer
$l{-}1$'s backward pass, the electronic unit finalizes factor
gradient accumulation for layer~$l$ and, if needed,
recomputes $\mathbf{X}_i$ for Eq.~(\ref{eq:grad_u}) by
executing a single forward sub-layer evaluation on a
temporarily available core (C4 or C5).
 
This staggered pipeline ensures that the optical cores and
Electronic units are rarely idle simultaneously. The key
enabler is that all weight transposes
$\mathbf{W}_{0,i}^\top$ are precomputed and stored in the
same weight memory used for $\mathbf{W}_{0,i}$ during the
forward phase, so the backward phase incurs no additional
tuning latency beyond the standard layer-by-layer streaming.

 \begin{figure} \vspace{-.3em}
 \centering
\includegraphics[width=0.42\textwidth]{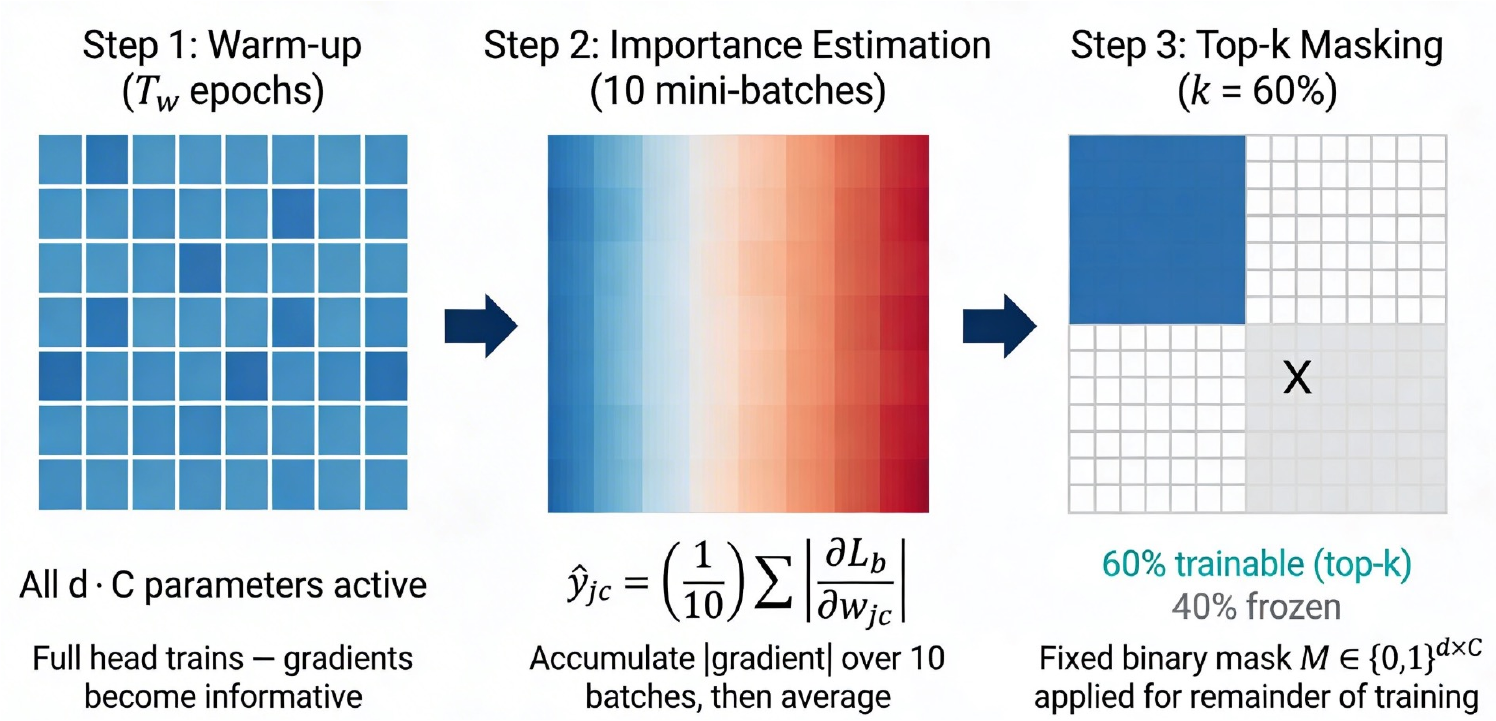} \vspace{-1em}
\caption{\footnotesize Gradient-Accumulated Sparse Head Training.}
\vspace{-3mm}
\label{ev2}
\end{figure}

\subsection{Gradient-Accumulated Sparse Head}
\label{sec:method_head}
 
The tensorized decomposition reduces backbone trainable
parameters to $\sim$8K for ViT-Base, but the task-specific
classification head
$\mathbf{W}_\text{head} \in \mathbb{R}^{d \times C}$
(where $C$ is the number of classes) requires $d \cdot C$
parameters. For $C{=}100$ (as in CIFAR-100 or many VTAB-1K
tasks), this is $768 \times 100 = 76{,}800$ parameters,
which is $9\times$ the backbone budget and dominates the
total trainable count. Applying low-rank factorization to
the head is ineffective: the classifier must map from a
shared $d$-dimensional representation to $C$
class-specific logits, and aggressive rank reduction blurs
inter-class boundaries, causing significant accuracy drops.
 
We propose an orthogonal approach: \emph{gradient-accumulated
sparse training} of the head. The method determines a fixed
sparsity mask early in training and applies it for the
remainder, proceeding in three steps.
 
\noindent\textbf{Step 1: Warm-up.}
We train the full head (all $d \cdot C$ parameters active) for
$T_w$ epochs, allowing the logit space to reach a coarse
alignment with the target task. During this phase, the
head parameters settle into a neighborhood where gradient
magnitudes become informative of long-term parameter
importance.
 
\noindent\textbf{Step 2: Importance estimation.}
After warm-up, we accumulate the absolute gradient magnitude
$|\partial\mathcal{L}/\partial w_{jc}|$ for each head
parameter $w_{jc}$ over $B_\text{acc}{=}10$ consecutive
mini-batches, then average:
\begin{equation}
\bar{g}_{jc} = \frac{1}{10}
  \sum_{b=1}^{10}
  \left|\frac{\partial\mathcal{L}_b}{\partial w_{jc}}\right|
\label{eq:importance}
\end{equation}
Accumulating over 10 batches provides sufficient
statistical stability to distinguish consistently important
parameters from those with transiently large gradients due
to mini-batch noise, while adding negligible overhead to the
training loop.
 
\noindent\textbf{Step 3: Top-$k$ masking.}
We rank head parameters by $\bar{g}_{jc}$ and construct a binary mask $\mathbf{M}$ that retains the top-$k$ fraction as trainable, while freezing the rest for the remainder of training. The update becomes $w_{jc} \leftarrow w_{jc} - \eta\,M_{jc}\,\partial\mathcal{L}/\partial w_{jc}$.

In practice, retaining $k{=}60\%$ yields negligible accuracy loss. The total trainable parameters are reduced to $\sim$54K (backbone + head). On the photonic hardware, the fixed mask means that 40\% of the head gradient computations and the corresponding optimizer state are eliminated for the entire training run after the brief warm-up and estimation phase,, reducing both SRAM footprint and backward compute cost.

\subsection{Photonic Noise Model for Training}
\label{sec:method_noise}
 
Existing SiPh accelerator studies~\cite{sunny2021crosslight,
sunny2021robin,morsali2024lightator,morsali2025optovit}
characterize MRR device noise exclusively for the inference
setting. In this work, we extend the noise model developed
for the Opto-ViT-v2 inference pipeline~\cite{morsali2025optovit}
to cover the full training loop, incorporating the same
device-level noise sources into both the forward and backward
optical passes.
 
\subsubsection{Noise Sources}
\label{sec:method_noise_sources}

Prior photonic noise modeling frameworks primarily treat device
non-idealities, including fabrication variation, thermal drift, and
laser noise, as \emph{additive or multiplicative stochastic
perturbations} with activation-dependent variance proxies derived
at the MAC or logit level~\cite{chen2026lightbound}. While such
models capture aggregate noise statistics, they assume
\emph{independent, zero-mean perturbations} and do not explicitly
model spatial thermal interactions or weight-dependent effects. In contrast, we focus on \textbf{thermo-optic crosstalk} as a
dominant source of structured, correlated noise in dense MRR
arrays. Our device-level characterization shows that heater-induced temperature variations propagate laterally across waveguides, introducing \emph{deterministic and spatially coupled resonance shifts} that depend on both \textbf{input optical power} and \textbf{inter-ring spacing}. Unlike prior stochastic models, this behavior cannot be captured by independent Gaussian noise. To address this, we model the effective resonance of each MRR as:
\begin{equation}
\lambda_i^{\text{act}}
=
\lambda_i^{\text{LUT}}
+
\underbrace{\alpha_i(P)\,\rho(d_i)}_{\text{thermal crosstalk}}
+
\mu_{r,i}
+
\epsilon_i,
\label{eq:thermal_model_updated}
\end{equation}
\vspace{-2.5mm}

where $\alpha_i(P)$ captures the \textbf{input power-dependent
thermal shift}, and $\rho(d_i)$ models \textbf{distance-dependent
heat diffusion} across neighboring rings. This formulation
explicitly captures \emph{weight-dependent noise}, since the
effective perturbation depends on the programmed detuning and
optical intensity at each MRR. Importantly, thermal drift also alters spectral spacing, leading to
\textbf{dynamic, input-dependent inter-channel crosstalk}: $\phi_{ij}^{\text{act}}=\frac{\delta_i^2}{(\lambda_i^{\text{act}} - \lambda_j^{\text{act}})^2 + \delta_i^2}$,
which couples neighboring channels through thermally shifted
resonances.

\subsubsection{Forward-Pass Noise}
\label{sec:method_noise_fwd}

Under the proposed thermo-optic cross-talk model, the optical
computation at sub-layer~$i$ is perturbed by structured,
weight-dependent noise:
\begin{equation}
\hat{\mathbf{Y}}_i^\text{opt}
  = \mathbf{X}_i\,\mathbf{W}_{0,i}
  + \bm{\epsilon}_i^\text{fwd}
\label{eq:fwd_noise}
\end{equation}
\vspace{-1em}

where $\bm{\epsilon}_i^\text{fwd}$ arises from thermally
induced, spatially correlated perturbations in the MRR
resonances and the resulting inter-channel crosstalk.
Unlike conventional i.i.d. noise models, this perturbation
is \emph{input- and weight-dependent}, since it is governed
by the local optical power and programmed de-tuning. The noisy optical output is combined with the clean electronic
factor contribution $s\,(\mathbf{X}_i \mathbf{U})\,\bm{\Sigma}_i
\mathbf{V}^\top$ and propagated through subsequent nonlinear
and residual operations, forming the input $\mathbf{X}_{i+1}$.
As a result, cached intermediates (e.g., $\mathbf{X}_i\mathbf{U}$)
implicitly accumulate forward noise across layers.

\subsubsection{Backward-Pass Noise}
\label{sec:method_noise_bwd}

The backward propagation
$\mathbf{G}_i\,\mathbf{W}_{0,i}^\top$ is executed on the same
optical cores and is subject to the same structured noise:
\begin{equation}
\widehat{\frac{\partial\mathcal{L}}{\partial\mathbf{X}_i}}
  \bigg|_\text{opt}
  = \mathbf{G}_i\,\mathbf{W}_{0,i}^\top
  + \bm{\epsilon}_i^\text{bwd}
\label{eq:bwd_noise}
\end{equation}
\vspace{-1em}

where $\bm{\epsilon}_i^\text{bwd}$ follows the same
weight- and input-dependent structure as
$\bm{\epsilon}_i^\text{fwd}$, but is statistically independent
due to different operands and operating conditions. Noise accumulates along the backward chain, such that gradients
at earlier layers are affected by both forward-pass corruption
(in activations) and backward-pass perturbations (in gradient
propagation).

\subsubsection{Impact on Factor Gradients}
\label{sec:method_noise_impact}

While factor gradient computations are performed electronically,
their inputs are corrupted: (i)~$\mathbf{X}_i \mathbf{U}$ carries
accumulated forward noise, and (ii)~$\mathbf{G}_i \mathbf{V}$
carries accumulated backward noise. Consequently, the electronic
unit computes exact updates on \emph{noisy operands}, which is
the primary mechanism through which structured photonic noise
influences learning. All noise parameters are calibrated from measured MRR behavior
and injected into both forward and backward optical computations
during training, enabling realistic evaluation of noise impact
on parameter-efficient fine-tuning.
\section{Experimental Results}
\label{sec:experiments}

\newcommand{\rot}[1]{\rotatebox{90}{#1}}

\begin{table}[t]
\centering
\scriptsize
\setlength{\tabcolsep}{2pt}
\begin{tabular}{l c ccccccc cccc c}
\toprule
& \#p 
& C100 & Cal & DTD & Flw & Pets & SVHN & Sun 
& Cam & Eur & Res & Ret 
& Avg \\
\midrule

Full & 85.9M 
& 68.9 & 87.7 & 64.3 & 97.2 & 86.9 & 87.4 & 38.8
& 79.7 & 95.7 & 84.2 & 73.9
& 78.6 \\

Adapter~\cite{houlsby2019parameter} & 217K 
& 69.2 & 90.1 & 68.0 & 98.8 & 89.9 & 82.8 & 54.3
& 84.0 & 94.9 & 81.9 & 75.5
& 80.9 \\

AdaptFormer~\cite{chen2022adaptformer} & 217K
& 70.8 & 91.2 & 70.5 & 99.1 & 90.9 & 86.6 & 54.8
& 83.0 & 95.8 & 84.4 & 76.3
& 82.1 \\

LoRA~\cite{hu2022lora} & 355K
& 67.1 & 91.4 & 69.4 & 98.8 & 90.4 & 85.3 & 54.0
& 84.9 & 95.3 & 84.4 & 73.6
& 81.3 \\

NOAH~\cite{zhang2022neural} & 421K
& 69.6 & 92.7 & 70.2 & 99.1 & 90.4 & 86.1 & 53.7
& 84.4 & 95.4 & 83.9 & 75.8
& 81.9 \\

FacT-TT$_4$~\cite{jie2023fact} & 68K
& 69.4 & 88.5 & 70.6 & 98.8 & 90.0 & 83.3 & 53.7
& 83.9 & 95.1 & 81.5 & 75.4
& 80.9 \\

FacT-TT$_{16}$~\cite{jie2023fact} & 96K
& 71.3 & 89.6 & 70.7 & 98.9 & 91.0 & 87.8 & 54.6
& 85.2 & 95.5 & 83.4 & 75.7
& \textbf{82.2} \\

\midrule

ViT-B (CL) & 60K 
& 43.4 & 85.7 & 64.7 & 97.1 & 86.6 & 76.4 & 49.3
& 79.3 & 90.3 & 76.4 & 74.3
& 74.9 \\

ViT-B$_4$ (FiT)  & \textbf{44K} 
& 67.0 & 88.1 & 68.1 & 98.6 & 89.2 & 86.5 & 51.5 
& 83.8 & 95.2 & 82.8 & 75.4 
& 80.6 \\

ViT-B$_{16}$ (FiT)  & 72K 
& 67.4 & 88.8 & 68.6 & 98.8 & 89.3 & 89.2 & 51.2 
& 86.1 & 95.8 & 84.1 & 76.1 
& 81.4 \\

\bottomrule
\end{tabular}
\caption{\footnotesize Results on the VTAB-1K benchmark under hardware-aware noise. `\#p' denotes the average number of trainable parameters, including both the backbone and the classification head. CL refers to classification-layer-only fine-tuning.}
\vspace{-2em}
\label{tab:vtab}
\end{table}

\begin{table*}[t]
\centering
\small
\setlength{\tabcolsep}{3pt}

\begin{tabular}{l c | cccc | cccc | cccc | cccc | cccc}
\toprule
Model & \#param 
& \multicolumn{4}{c}{FGVCAircraft}
& \multicolumn{4}{c}{OxfordPets}
& \multicolumn{4}{c}{Food101}
& \multicolumn{4}{c}{StanfordCars}
& \multicolumn{4}{c}{Flowers102} \\

\cmidrule(lr){3-6} \cmidrule(lr){7-10} \cmidrule(lr){11-14} \cmidrule(lr){15-18} \cmidrule(lr){19-22}

& 
& 2 & 4 & 8 & 16
& 2 & 4 & 8 & 16
& 2 & 4 & 8 & 16
& 2 & 4 & 8 & 16
& 2 & 4 & 8 & 16 \\

\midrule

Adapter~\cite{houlsby2019parameter} & 239K
& 11.9 & 21.5 & 32.1 & 45.0
& 76.3 & 85.3 & 87.5 & 87.8
& 48.6 & 60.0 & 66.2 & 71.1
& 12.3 & 23.8 & 40.0 & 60.0
& 96.4 & 97.4 & 98.9 & 99.3\\

AdaptFormer~\cite{chen2022adaptformer} & 239K
& 12.0 & 21.6 & 34.2 & 48.9 
& 74.9 & 83.9 & 86.3 & 87.5
& 48.3 & 59.8 & 66.3 & 71.4
& 12.3 & 26.5 & 48.5 & 65.2
& 96.4 & 97.3 & 98.9 & 99.3\\

LoRA~\cite{hu2022lora} & 377K
& 12.1 & 21.8 & 34.5 & 48.7 
& 76.2 & 85.6 & 86.4 & 87.4
& 48.6 & 59.5 & 66.1 & 71.4
& 12.3 & 26.4 & 48.2 & 67.8
& 96.5 & 97.3 & 98.7 & 99.4\\

NOAH~\cite{zhang2022neural} & 417K
& 12.3 & 22.2 & 34.5 & 49.2
& 78.2 & 84.9 & 87.4 & 89.4
& 51.5 & 64.1 & 71.9 & 74.7
& 12.2 & 25.5 & 48.5 & 68.2
& 97.0 & 97.5 & 98.9 & 99.5\\

FacT-TT$_{16}$~\cite{jie2023fact} & 143K
& 15.7 & 24.5 & 36.8 & 50.8
& 82.2 & 87.5 & 89.8 & 90.0
& 53.8 & 64.5 & 71.5 & 75.0
& 16.2 & 29.8 & 52.5 & 74.3
& 90.0 & 97.0 & 98.9 & 99.3 \\

\midrule

ViT-B$_4$ (FiT) & 58K
& 17.2 & 27.1 & 37.6 & 51.1
& 80.8 & 86.3 & 90.1 & 91.0
& 47.8 & 61.4 & 68.8 & 74.1
& 13.9 & 23.6 & 40.7 & 57.8
& 97.1 & 98.1 & 98.8 & 98.9 \\

ViT-B$_{16}$ (FiT) & 86K
& 17.2 & 28.3 & 40.4 & 56.1
& 82.4 & 86.8 & 89.4 & 90.0
& 47.8 & 59.8 & 66.8 & 72.6
& 14.1 & 26.7 & 50.0 & 69.5
& 97.6 & 97.8 & 98.7 & 98.9 \\

\bottomrule
\end{tabular}

\vspace{0.5em}
\caption{
Few-shot results from 2 to 16 shots on FGVC benchmarks under hardware-aware noise. `\#param' denotes the average number of trainable parameters, including both the backbone and the classification head.
FacT-TT rank-16 results are obtained under noise-free conditions, while FiT method evaluates the same setting with a 40\% sparse head under hardware-induced noise.
} \vspace{-3em}
\label{tab:fgvc}
\end{table*}

\begin{figure}[t] \vspace{-1em}
\centering
\includegraphics [width=0.85\linewidth]{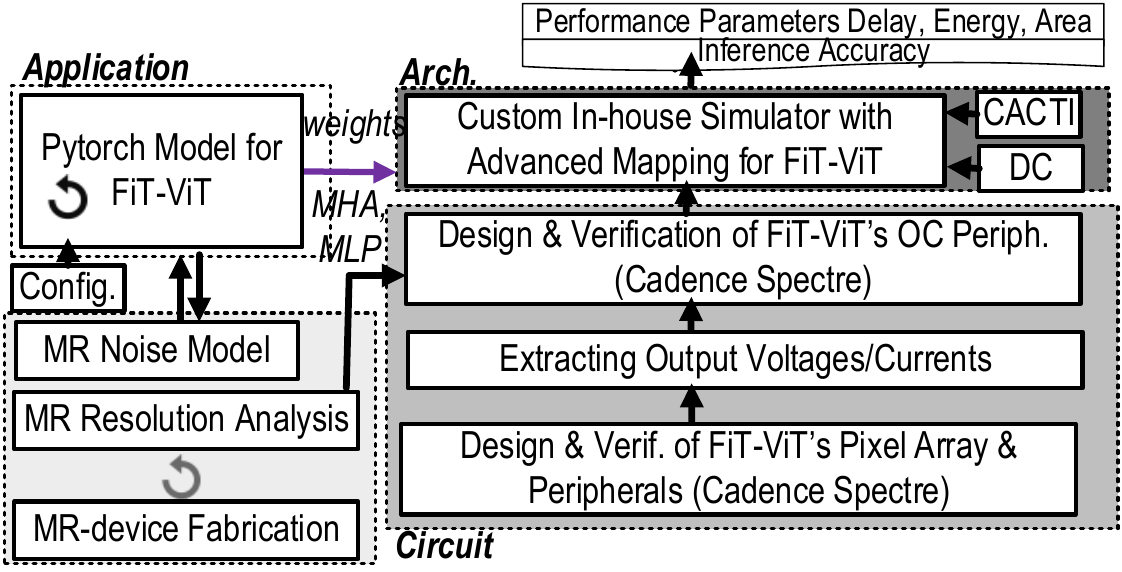}
\vspace{-1.1em}
\caption{Proposed bottom-up evaluation framework.}
\vspace{-2em}
\label{framework}
\end{figure}

\subsection{Evaluation Framework} We develop a comprehensive bottom-up framework spanning device, circuit, architecture, and application levels (Fig.~\ref{framework}).

\noindent\textbf{At the device level}, MRRs were fabricated and calibrated to achieve 8-bit precision, with over 200 identical instances integrated on a $10\times10~\mathrm{mm}^2$ chip (Fig.~2(c)) to characterize process variations. Thermo-optic behavior and thermal crosstalk were analyzed using coupled Ansys Lumerical HEAT and FDE simulations. A silicon waveguide ($500,\mathrm{nm} \times 220,\mathrm{nm}$) with a heater positioned $2,\mu\mathrm{m}$ above was modeled, with a secondary waveguide placed at separations of $2$–$10,\mu\mathrm{m}$ to study spatial thermal coupling. Sweeping heater power ($0$–$6,\mathrm{mW}$), we extracted temperature distributions, effective index changes (using a thermo-optic coefficient of $1.8 \times 10^{-4},\mathrm{K}^{-1}$), and resulting phase and wavelength shifts. Thermal crosstalk was quantified via the relative temperature rise in the secondary waveguide. These measurements were co-simulated with interface CMOS circuits in Cadence Spectre. 

\noindent\textbf{At the circuit level}, pixel arrays and peripherals were implemented using the 45nm NCSU PDK~\cite{NCSU_PDK}, enabling extraction of input–output characteristics. Building on this, all Opto-ViT-v2 components (excluding memory) were designed and validated using Cadence Spectre and Synopsys Design Compiler~\cite{DC}, and integrated into our in-house simulator for holistic latency and energy evaluation across configurations.

\noindent\textbf{At the application level}, we evaluate our method on the Visual Task Adaptation Benchmark (VTAB-1K)~\cite{zhai2019vtab} and several fine-grained visual classification (FGVC) datasets. 
VTAB-1K comprises 19 classification tasks with 1,000 training samples each, covering natural, specialized, and structured domains, and is widely used for evaluating parameter-efficient transfer learning under distribution shifts. We further evaluate on five FGVC benchmarks: Food-101~\cite{bossard2014food101}, Stanford Cars~\cite{krause2013cars}, Oxford Flowers-102~\cite{nilsback2008flowers}, FGVC-Aircraft~\cite{maji2013fgvcaircraft}, and Oxford-IIIT Pets~\cite{parkhi2012catsdogs}, following the standard few-shot protocol. In order to report the top-1 accuracy of them, all models are initialized from an ImageNet-21K pretrained ViT-Base, with backbone weights quantized to 8-bit and kept frozen; only the proposed FacT parameters and classification head are updated.

To evaluate hardware-aware robustness, we inject structured photonic noise during both forward and backward propagation. Following the thermo-optic crosstalk model in Section~3.3, weight perturbations are obtained via a lookup-table (LUT) that maps programmed weights to their \emph{input- and weight-dependent} resonance shifts. This captures the deterministic component of thermal drift arising from optical power and spatial coupling. On top of this, we model residual stochastic variations as zero-mean Gaussian noise with standard deviation $\sigma = 0.12$, consistent with our post-trimming device-level measurements. Input activations are similarly perturbed to reflect optical intensity fluctuations, using Gaussian noise with $\sigma = 0.03$. These perturbations are propagated through both forward and backward passes, ensuring that gradients are computed under the same structured, hardware-consistent noise conditions. For fine-tuning, we use AdamW with a learning rate of $5\times10^{-3}$ and weight decay of $10^{-4}$. A cosine schedule is applied with a 5-epoch warmup at $10^{-6}$, followed by 95 epochs of training. 

\subsection{Opto-ViT-v2 Results Analysis} 

Table~\ref{tab:vtab} and Table~\ref{tab:fgvc} present the performance of Opto-ViT-v2 on VTAB-1K and FGVC benchmarks under hardware-aware photonic noise. The results show that Opto-ViT-v2 maintains strong adaptation capability while operating under strict parameter and hardware constraints.

\noindent\textbf{Accuracy analysis.}
Opto-ViT-v2 with rank-16 achieves competitive performance across both VTAB-1K and FGVC tasks, with accuracy comparable to state-of-the-art PEFT methods. This indicates that the proposed tensorized low-rank adaptation retains most of the representational capacity of full fine-tuning, even under structured photonic noise.  The rank-4 configuration shows a moderate drop in accuracy, but still consistently outperforms the classification-layer-only baseline. Nevertheless, the results demonstrate that a small number of shared factors is sufficient to capture task-specific adaptation because it still have comparable result under the tested VTAB-1K datasets

On FGVC benchmarks, Opto-ViT-v2 performs particularly well in low-shot regimes, retaining the few-shot accuracy compared to FacT-TT's result. It suggests that the structured low-rank parameterization together with sparse updates introduces an implicit regularization effect. This helps prevent overfitting when training data is limited, while still enabling effective adaptation.

\noindent\textbf{Parameter efficiency and robustness.}
Opto-ViT-v2 achieves these results with substantially fewer trainable parameters than prior methods. While existing approaches require 200K–300K parameters, Opto-ViT-v2 operates with only 44K–72K parameters, corresponding to a $3\times$–$8\times$ reduction. 

Importantly, the total training cost is not dominated by the backbone alone, but also by the classification head. In many VTAB and FGVC tasks, the head accounts for the majority of trainable parameters. The sparse head of Opto-ViT-v2 helps remove low-importance parameters and reduces the head size by approximately 40\% with minimal impact on accuracy. This significantly lowers both memory footprint and update cost, making the method more suitable for on-chip training. Meanwhile, Opto-ViT-v2 remains robust under hardware noise. By constraining updates to a low-dimensional factor space and eliminating unnecessary head updates, the method reduces sensitivity to noisy gradients and activations. As a result, Opto-ViT-v2 maintains accuracy close to noise-free baselines despite the presence of structured photonic perturbations.

\noindent\textbf{Sparse Head and Noise Ablation Study.}
The effect of sparse-head design and hardware noise is further validated by Table~\ref{tab:ablation}. Increasing the retaining ratio from 0.4 to 0.6 improves accuracy, while further increasing to 0.8 provides only marginal gains, indicating that most useful parameters are concentrated in a subset of the head. This confirms that the proposed 0.6 retaining ratio strategy effectively reduces parameter count.

As fabrication noise increases, accuracy degrades across all tasks, with more pronounced drops on complex datasets. This reflects the impact of noise on both forward activations and backward gradients. Nevertheless, Opto-ViT-v2 degrades gracefully, demonstrating that the proposed low-rank and sparse-update design improves stability under realistic hardware conditions.

\begin{table}[t] 
\centering
\footnotesize
\setlength{\tabcolsep}{2pt}

\resizebox{0.98\columnwidth}{!}{%
\begin{tabular}{c c c c c c c c }
\toprule
\multicolumn{8}{c}{\textbf{Effect of Head Retaining Ratio}} \\
\midrule
Retaining ratio & \# Head Params
& CIFAR100 & Pets & RESISC45 & Camelyon & CLEVR-Count & KITTI  \\
\midrule
0.4 & 10K & 65.5 & 89.2 & 80.6 & 82.9 & 74.4 & 77.9  \\
0.6 & 15K & 67.0 & 89.2 & 82.8 & 83.8 & 73.5 & 78.2  \\
0.8 & 20K & 67.9 & 90.0 & 83.7 & 83.6 & 73.4 & 78.3  \\
\midrule
\multicolumn{8}{c}{\textbf{Effect of Fabrication Noise $\sigma_{\mathrm{fab}}$}} \\
\midrule
$\sigma_{\mathrm{fab}}$ & \# Head Params
& CIFAR100 & Pets & RESISC45 & Camelyon & CLEVR-Count & KITTI  \\
\midrule
0.1 & \multirow{3}{*}{15K}
& 67.0 & 89.2 & 82.8 & 83.8 & 73.5 & 78.2 \\
0.2 & 
& 64.2 & 88.5 & 81.2 & 82.4 & 69.6 & 77.4  \\
0.3 & 
& 58.1 & 85.5 & 78.3 & 81.8 & 62.0 & 76.1  \\
\bottomrule
\end{tabular}%
}

\caption{
Ablation study on head retaining ratio and fabrication noise in serveral Vtab1K datasets.
For sparse head analysis, $\sigma_{\mathrm{fab}}$ is fixed at 0.1.
For noise analysis, retaining ratio is fixed at 0.6.
``\# Head Params'' denotes the average number of trainable classifier-head parameters.
} 
\vspace{-2em}
\label{tab:ablation}
\end{table}

\subsection{Energy and Latency Analysis}

We evaluate one full training epoch for a ViT-Base-style backbone under the VTAB-1K setting using the proposed customized evaluation framework designed to capture forward propagation, backward propagation, gradient/factor computation, activation caching or recomputation, and parameter update/write-back overheads.
We assume a ViT-Base-like configuration with embedding dimension $d=768$, depth $L=12$, and $12L=144$ trainable backbone matrices after tensorization into the unified sub-layer representation used by FacT-style factorization. For VTAB-1K, we assume 1000 training samples per task and task-specific classifier dimensions $C$, while keeping the same backbone and hardware configuration across tasks. The number of optimization steps per epoch is therefore $\lceil 1000/B \rceil$, where $B$ is the batch size.

For each model, we decompose epoch cost into the following phases:
\begin{equation}
E_{\mathrm{epoch}} =
E_{\mathrm{fwd}} + E_{\mathrm{bwd}} + E_{\mathrm{grad/update}} + E_{\mathrm{update/tune}} + E_{\mathrm{mem/recompute}} + E_{\mathrm{head}},
\end{equation}
\begin{equation}
T_{\mathrm{epoch}} =
T_{\mathrm{fwd}} + T_{\mathrm{bwd}} + T_{\mathrm{grad/update}} + T_{\mathrm{update/tune}} + T_{\mathrm{mem/recompute}} + T_{\mathrm{head}}.
\end{equation}
These terms are accumulated over all optimization steps in one epoch.

\noindent\textbf{Comapred Architectures.}
We evaluate four photonic training architectures along with an FPGA baseline, differing in how matrix multiplications and parameter updates are mapped across optical and electronic domains. In the \textbf{[O:E]} model, forward and backward matrix multiplications are executed optically, while full-rank gradient computation and weight updates are performed electronically, avoiding optical write-back but incurring large gradient and memory costs. The proposed \textbf{[O:E$_L$]} model extends this design by adopting a low-rank factorization, where only compact factors are trained electronically, significantly reducing gradient dimensionality, activation storage, and update overhead. In contrast, the \textbf{[O:O]} model represents a fully optical training approach in which both propagation and full-rank weight updates are implemented in optical cores, requiring extensive microring tuning and write-back operations, leading to high update energy and latency. The \textbf{[O:O$_L$]} model combines optical execution with low-rank training, keeping pretrained weights fixed on optical cores while updating only a small set of low-rank parameters. Finally, we include a {Xilinx VCK190 FPGA baseline}, using its reported energy efficiency (1.42 KFPS/W) for total energy comparison.

\begin{figure}[t] \vspace{-1.5em}
 \centering
\includegraphics[width=0.5\textwidth]{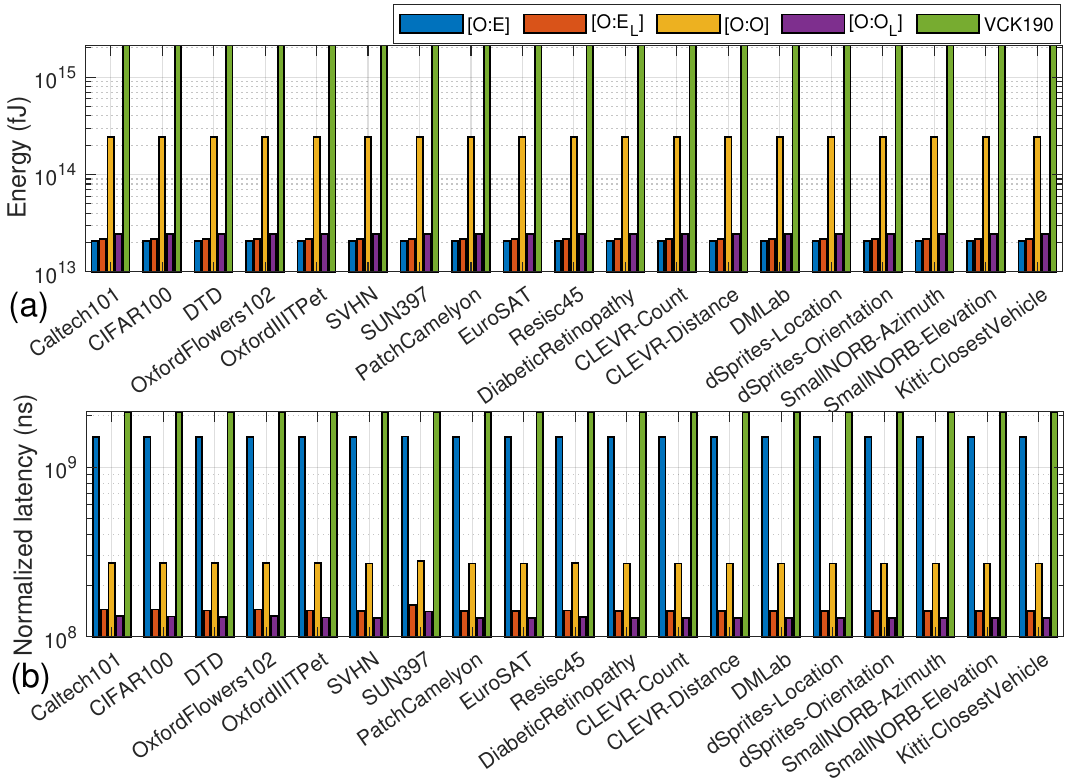} \vspace{-2em}
\caption{ Per-task (a) total epoch energy and (b) Normalized total latency.}
\vspace{-2em}
\label{ev1}
\end{figure}

\noindent\textbf{Results.}
As shown in Fig.~\ref{ev1}, the fully optical model \textbf{[O:O]} incurs the highest cost due to updating the entire backbone in the optical domain. The number of tunable parameters scales as $12Ld^2 \approx 8.5\times10^7$, resulting in $\sim1.7\times10^8$ optical write operations per step (including $W$ and $W^\top$), making update/tuning the dominant energy and latency bottleneck. In contrast, the low-rank design \textbf{[O:O$_L$]} reduces trainable parameters to $2dr + 12Lr^2 = 8448$ ($d{=}768$, $r{=}4$), i.e., $\sim10^4\times$ fewer parameters, and correspondingly reduces programming events to $\sim1.7\times10^4$, yielding orders-of-magnitude savings in update energy and latency. A similar trend holds for electronic-update variants. While \textbf{[O:E]} avoids optical write-back, it still incurs full-rank gradient computation and large activation storage. The proposed \textbf{[O:E$_L$]} reduces gradient complexity from $\mathcal{O}(d^2)$ to $\mathcal{O}(dr)$ and shrinks cached activations by $\sim192\times$ (from $2.17\times10^7$ to $1.13\times10^5$ elements), leading to consistently lower energy and latency. Across VTAB tasks, variation is small since costs are dominated by the shared ViT backbone. Overall, these results show that weight updates—not forward/backward passes—are the primary bottleneck in photonic training, and that low-rank factorization is key to scalable efficiency.

The average epoch energy breakdown in Fig.~\ref{ev2}(a) shows that forward and backward propagation dominate due to shared optical execution. However, the \textbf{Update/Tuning} phase exhibits the largest variation. The fully optical \textbf{[O:O]} design incurs the highest update energy from $\sim1.7\times10^8$ programming events per step, while \textbf{[O:O$_L$]} reduces this to $\sim1.7\times10^4$, achieving an $\sim10^4\times$ reduction. Electronic-update models eliminate optical tuning, with \textbf{[O:E$_L$]} further lowering gradient energy via low-rank factorization. Memory/recomputation energy is also reduced due to a $\sim192\times$ decrease in cached activations. Latency trends in Fig.~\ref{ev2}(b) mirror energy, with forward/backward passes dominating the baseline. However, \textbf{Update/Tuning} dominates latency in full optical training. \textbf{[O:O]} suffers from sequential programming scaling with $\mathcal{O}(d^2)$, while \textbf{[O:O$_L$]} reduces this to only $8448$ parameters, yielding orders-of-magnitude lower latency. Electronic-update models \textbf{[O:E]} and \textbf{[O:E$_L$]} avoid optical programming entirely, achieving the lowest update latency, with \textbf{[O:E$_L$]} benefiting from reduced gradient complexity. From a resource perspective, optical computation dominates forward/backward latency across all models. However, \textbf{Update/Tuning} is the primary bottleneck in \textbf{[O:O]} due to $\sim10^8$ writes, while \textbf{[O:O$_L$]} reduces this by $\sim10^4\times$. Electronic computation remains modest but increases in \textbf{[O:E]} and \textbf{[O:E$_L$]} due to gradient updates. Memory latency is also reduced in low-rank models, benefiting from the $\sim192\times$ reduction in activation storage.

 \begin{figure}[t] 
 \centering
\includegraphics[width=0.5\textwidth]{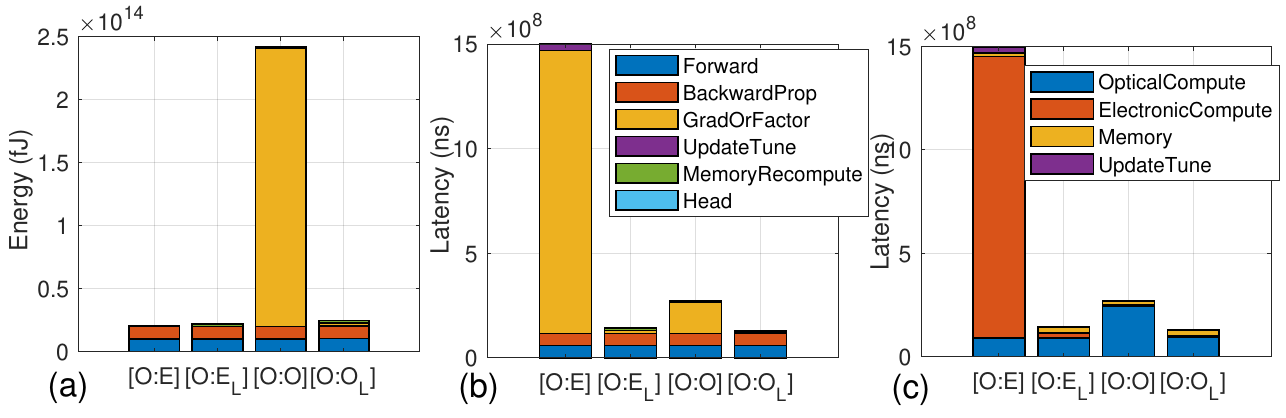} \vspace{-2em}
\caption{Average epoch (a) phase and (b) Latency by phase and (c) Latency by resource.}
\vspace{-2em}
\label{ev2}
\end{figure}

\section{Conclusion}
\label{sec:conclusion}

This work presents Opto-ViT-v2, a hardware–algorithm co-designed
framework for parameter-efficient on-chip fine-tuning on silicon-
photonic ViT accelerators. By combining tensorized
low-rank adaptation with a gradient-accumulated sparse head, FiT-
ViT significantly reduces training parameters while preserving
adaptation capability. Opto-ViT-v2 achieves competitive accuracy with fewer parameters and
remains robust under structured photonic noise, while reducing
activation storage and eliminating costly weight write-back. This
demonstrates that low-dimensional and selective updates enable
stable and efficient on-chip learning on photonic hardware.
 
 


\end{document}